\documentclass[11pt,twoside]{article}
\usepackage{asp2010}

\resetcounters

\begin{document}

\title{Weighing Betelgeuse: Measuring the mass of $\alpha$ Orionis from stellar limb-darkening}
\author{Hilding R. Neilson$^1$, John B. Lester$^{2,3}$, and Xavier Haubois$^4$
\affil{$^1$Argelander-Institut f\"{u}r Astronomie, Auf dem H\"{u}gel 71, D-53121 Bonn, Germany}
\affil{$^2$Department of Physical and Chemical Sciences, University of Toronto Mississauga}
\affil{$^3$Department of Astronomy and Astrophysics, University of Toronto}
\affil{$^4$Instituto de Astronomia, Geof\'{i}sica e Ci\^{e}ncias Atmosf\'{e}ricas,
Universidade de S\~{a}o Paulo, Rua do Mat\~{a}o 1226, Cidade
Universit\'{a}ria, S\~{a}o Paulo, SP 05508-900, Brazil}
}

\begin{abstract}
Stellar limb-darkening is an important tool for constraining the properties of a stellar atmosphere. We present a novel method for relating the fundamental stellar parameters mass and radius to limb-darkening laws using grids of spherical model stellar atmospheres. This method is applied to interferometric observations of the red supergiant Betelgeuse, where an unique measure of the stellar mass is determined. 
\end{abstract}
\section{Introduction}
The M2Iab star Betelgeuse is one of the brightest stars in the night, and one of the most studied. However, for all that is known about the star, there are still a number of mysteries.  One such mystery is the mass of Betelgeuse.  As a star without a binary companion, there exists no direct method to measure its mass, a mass estimate is possible only using theoretical modeling. \cite{Dolan2008} fit stellar evolution models to measured values of the stellar radius, effective temperature and mass-loss rate to find a stellar mass $M = 21\pm 2~M_\odot$.  Furthermore, \cite{Lobel2000} used detailed sprectral synthesis to derive the surface gravity of $\log g = -0.5$ for Betelgeuse.  Coupled with the radius assumed by \cite{Dolan2008}, that gravity implies a mass $M\approx 9.5~M_\odot$.  Clearly, these two estimates do not agree.  

The purpose of this work is to present a new measurement of the mass of Betelgeuse, using a novel method based on observations of the star's intensity profile from narrow H-band interferometry \citep{Haubois2009}, the properties of flux-conserving limb-darkening laws, and spherically-symmetric model stellar atmospheres. 

\section{Model Stellar Atmospheres}
We compute a grid of model stellar atmospheres using the  \textsc{SAtlas} code developed by \cite{Lester2008}, which is a spherical version of the \textsc{Atlas} code developed by \cite{Kurucz1979}. The code assumes local thermodynamic equilibrium and spherically-symmetric hydrostatic equilibrium and radiative transfer.  The code has been tested by comparison to spherically-symmetric \textsc{Phoenix} \citep{Hauschildt1999} and \textsc{Marcs} \citep{Gustafsson2008} models.  Computed intensity profiles have been used to fit K-band interferometric observations \citep{Neilson2008}, agreeing with previous results \citep{Wittkowski2004}.  

The grid is computed in three dimensions, varying effective temperature, gravity and stellar mass.  The effective temperature range is $3000$ - $8000~$K in steps of $200~$K, the gravity range is from $\log g = -1$ - $+3$ in cgs units in steps of  $0.25$, and the mass range is $M = 2.5$ - $20~M_\odot$ in steps of $2.5~M_\odot$.  There are approximately $2000$ models in this grid, although not all combinations of the fundamental parameters produced converged models.  Model intensity profiles are computed as a function of wavelength and $\mu$, where $\mu$ is the cosine of the angle formed by the line-of-sight point on a stellar disk and the center of the disk.  Each profile is computed for 1000 equally spaced $\mu$-points.

\section{Limb-darkening law}
In this work we consider the limb-darkening law
\begin{equation}
\frac{I}{2\mathcal{H}} = 1 - A - B + \frac{3}{2}A\mu + \frac{5}{4}B\sqrt{\mu},
\end{equation}
where $\mathcal{H}$ is the Eddington flux.  This law has two important properties that differ from other types of laws.  The first property is that the law conserves stellar flux by definition. The second property is of interest for this work.  \cite{Fields2003} found that limb-darkening fits to intensity profiles from plane-parallel model atmospheres predicted relations that {\it{all}} intersected at a fixed point, $\mu_1$, with the same normalized intensity. The authors compared these profiles to limb-darkening relations from microlensing observations, the observations did not agree.  \cite{Neilson2011a} showed that this fixed point occurs because the limb-darkening coefficients $A$, and $B$, are correlated such that $A \propto \alpha B$.  The slope of the correlation is a function of the ratio $\eta \equiv \int I(\mu)\sqrt{\mu}d\mu / J$, where $J$ is the mean intensity.  This ratio is comparable to the Eddington factor used in modelling stellar atmospheres $f = K/J$, where $K$ is the second moment of the intensity \citep{Mihalas1978}.  In plane-parallel model stellar atmospheres, the Eddington factor is approximately constant, and similarly $\eta$ is also constant.

\begin{figure}[t]
\begin{center}
\includegraphics[width=0.6\textwidth]{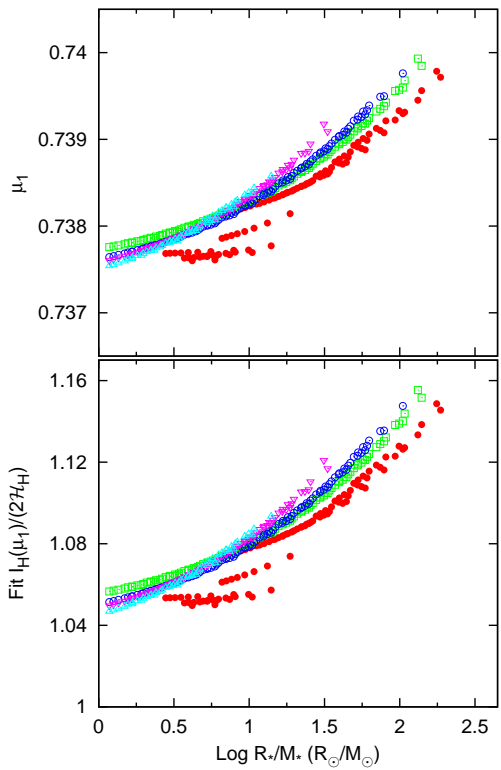}
\end{center}
\caption{The fixed point $\mu_1$ and normalized narrow H-band fit intensity at the fixed point $I_H(\mu_1)/2\mathcal{H}$ as a function of atmospheric extension parameterized as $R_*/M_*$.  Red filled circles correspond to $T_{\rm{eff}} = 3000~$K models, green open squares 4000 K, blue open circles 5000 K, magenta downward pointing triangles 6000 K and pale blue upward pointing triangles are 7000 K. }\label{f1}
\end{figure}

In spherically symmetric model stellar atmospheres, the Eddington factor and $\eta$ are not constant.  This is discussed in greater detail in \cite{Neilson2011a}. \cite{Neilson2011b} discovered that the variation of $\eta$, hence $\alpha$, and hence the fixed point and intensity at the fixed point is correlated to the amount of extension of the stellar atmosphere.  Furthermore, atmospheric extension can be shown to be a function of the ratio of the stellar radius and mass, $R_*/M_*$.  Therefore, if one can measure the limb-darkening coefficients for this law then one has a measure of the atmospheric extension and the ratio of the radius and mass. The relation between the values of $R_*/M_*$ and the fixed point $\mu_1$ and the intensity at the fixed point  $I(\mu_1)/2\mathcal{H}$ is shown in Fig.~\ref{f1} for our models with effective temperature $T_{\rm{eff}} = 3000,4000,5000,6000$, and $7000$~K. The fixed point and intensity are functions of $R_*/M_*$, such that
\begin{eqnarray}
\mu_1 = C_\mu \left(\log R_*/M_* \right)^2 + D_\mu, \\
\frac{I(\mu_1)}{2\mathcal{H}} = C_I \left(\log R_*/M_* \right)^2 + D_I,
\end{eqnarray} 
for some given effective temperatures.
\section{Method}
\begin{figure}[t]
\begin{center}
\includegraphics[width=\textwidth]{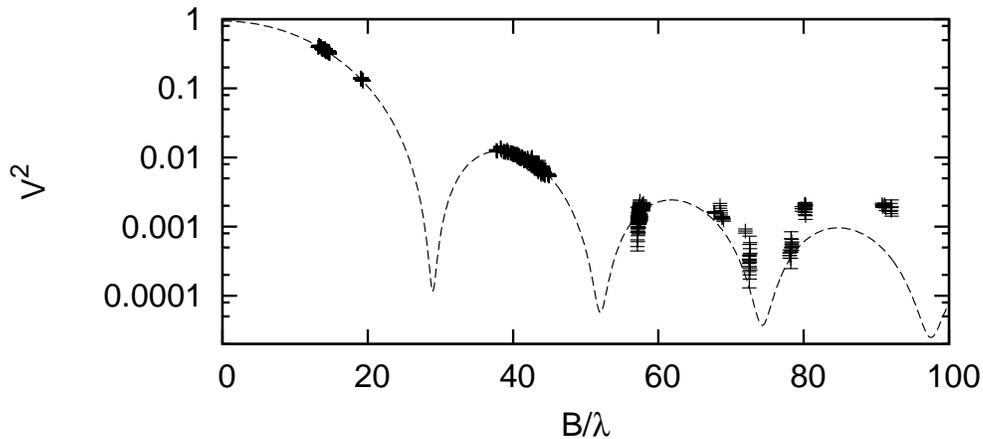}
\end{center}
\caption{The observed squared visibilities from \cite{Haubois2009} are shown as crosses as a function of baseline divided by effective wavelength $\lambda_0 = 1.64~\mu m$ in cycles per arcsecond, along with the predicted squared visibility, shown as a dashed line, for a $T_{\rm{eff}} = 3600~$K, $\log g = 0$, and $M=12.5~M_\odot$ intensity profile.}\label{f0}
\end{figure}

The apparent existence of the fixed point in the chosen limb-darkening law combined with interferometric observations from \cite{Haubois2009} suggests a method to measure the mass of Betelgeuse.  The first step is to determine the stellar radius for Betelgeuse.  We fit the interferometric observations, shown in Fig.~\ref{f0} along with the predicted visibilities from a stellar atmosphere model, to using narrow H-band intensity profiles from the model atmospheres to get a best-fit angular diameter.  The angular diameter can be combined with the measured distance, $d = 197\pm 45~$pc \citep{Harper2008},  and the bolometric flux $F_{\rm{bol}} = (111.67\pm6.49)\times 10^{-13}~W~$cm$^{-2}$ \citep{Perrin2004} of Betelgeuse to predict the stellar radius, and effective temperature.

The second step is to fit interferometric observations again, but with the limb-darkening law to measure the coefficients, $A$, and $B$.  Using the correlations from Eq.~2 and 3, with the limb-darkening law Eq.~1, we have three equations and three unknowns that can be solved to determine $R_*/M_*$.  Combining this result with the predicted radius, then we can measure the mass of Betelgeuse.  It should be noted that the coefficients, $A,B$ are degenerate functions of spherical extension because they are both functions of $\int I\sqrt{\mu}d\mu /(2\mathcal{H})$ and $J/(2\mathcal{H})$.  Therefore, one should not directly compare the values of the coefficients to the values of $R_*/M_*$.

\section{Results}
We fit the interferometric observations in the same manner as \cite{Haubois2009}, who fit intensity profiles to the observations in the first two lobes of the visibility curve, even though the visibility curve is observed to fourth lobe. The authors noted that the visibility curve is affected by asymmetries in the photosphere, so it should be ignored. We directly fit intensity profiles from the model atmosphere grid to the first two lobes of the interferometric observations, and measure the angular diameter of Betelgeuse to be $\theta = 44.93\pm 0.15~$mas, agreeing with \cite{Haubois2009}.  The combination of this angular diameter and bolometric flux and distance yields an effective temperature $T_{\rm{eff}} = 3590\pm55~$K and $R = 955\pm 217~R_\odot$, respectively.
\begin{figure}[t]
\begin{center}
\includegraphics[width=\textwidth]{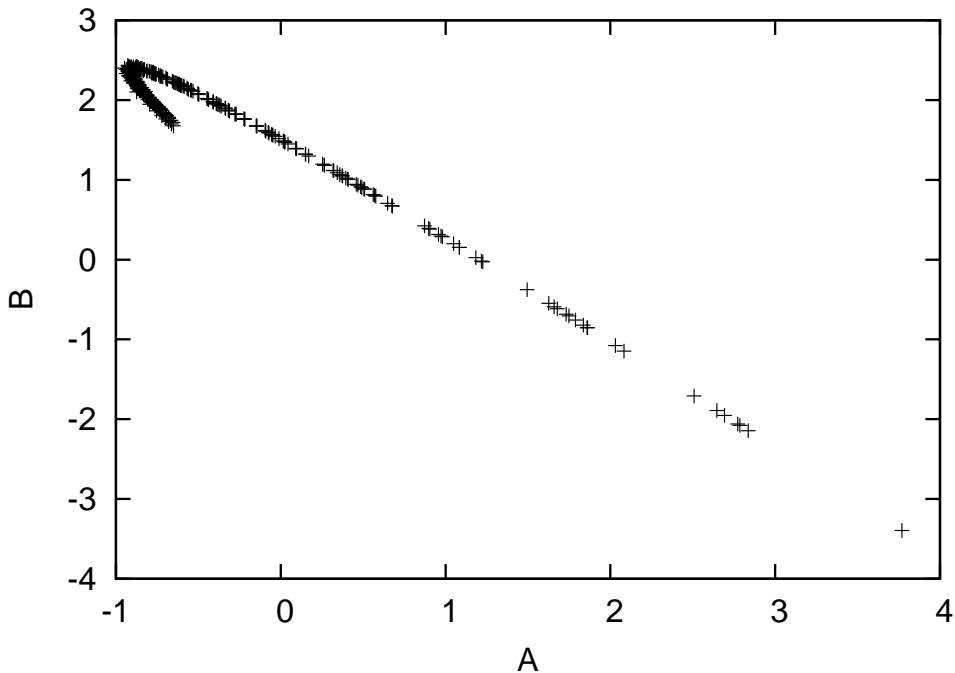}
\end{center}
\caption{Correlation between limb-darkening coefficients $A$ and $B$ from spherically-symmetric model stellar atmospheres with effective temperatures $T_{\rm{eff}} = 3400$ - $3800~$K.}\label{f2}
\end{figure}

The next step is to fit the limb-darkening law Eq.~1 to the interferometry data.  We try to determine limb-darkening coefficients from the first two lobes of the visibility curve, but the resulting limb-darkening fit predicts large intensities near the limb, $I(0)/2\mathcal{H} > 0.7$, which is greater than one-half of the central intensity.  This suggests that the second lobe of the visibility curve also contains light from circumstellar excess.  Therefore, we attempt to fit the limb-darkening law to the first lobe only, even though the first lobe of the visibility curve is a weak measure of the intensity profile.  We also consider the predicted limb-darkening coefficients for models with effective temperatures ranging from $3400$~K to $3800~$K.  In Fig.~\ref{f2}, we plot the limb-darkening coefficients  from model atmospheres in that temperature range.  In the range of $A = 0$ - $3$ there is a tight correlation between the coefficients of the form $B =-1.27A+1.50$ .  Using this correlation, we reduce the number of degrees-of-freedom.

When we fit the observed first lobe of the visibility curve for the limb-darkening coefficients, we find $A = 1.36\pm0.33$, which leads to $B = -0.22\pm 0.43$.  Using these limb-darkening coefficients with the correlations, Eq.~2, and 3, where $C_\mu = 4.338\times 10^{-4}$, $D_\mu = 0.73704$, $C_I = 0.01966$, and $D_I=1.0601$, we find a value of $R_*/M_* = 82.17^{+13.32}_{-11.51}R_\odot/M_\odot$ for Betelgeuse, hence for $R = 955\pm217~R_\odot$ we get $M = 11.6^{+5.0}_{-3.9}~M_\odot$.  Using the mass and radius, we predict the gravity to be $\log g = -0.45^{+0.17}_{-0.16}$.

As a check, we compare these results to limb-darkening fits to model atmospheres, the ''nearest`` model to the predicted effective temperature, gravity and mass is a $T_{\rm{eff}} = 3600~$K, $\log g = -0.5$, and $M = 10~M_\odot$ model atmosphere.  Its limb-darkening coefficients are $A = 1.63$ and $B = -0.55$ with a ratio $R_*/M_*  = 93.32~R_\odot/M_\odot$.  This ratio of $R_*/M_*$ agrees with the predicted ratio for Betelgeuse.   

\section{Summary}
In this work, we use a flux-conserving limb-darkening law, spherically-symmetric model stellar atmospheres computed with the \textsc{SAtlas} code, and interferometric observations to probe the fundamental parameters of Betelgeuse.  This novel method computed one of the first measurements of the mass of Betelgeuse.  The predicted mass is less than that determined by stellar evolution calculations \citep{Dolan2008}, but it is consistent with the gravity suggested by \cite{Lobel2000}.

The predicted mass is still very uncertain, the two main sources of uncertainties are the distance to Betelgeuse and the circumstellar flux excess apparent in the interferometric observations.  In spite of the large uncertainties, the mass prediction is precise enough to agree with \cite{Smith2009} that Betelgeuse will explode as a Type IIP supernova, and may even act as a constraint for stellar evolution models and input physics, such as convective core overshooting.

It is also worth noting that interferometry at shorter wavelengths may be a more effective constraint for our method.  At shorter wavelengths, interferometry has greater angular resolution, as well as avoiding circumstellar emission in the infrared.

\acknowledgements HRN is grateful for funding from the Alexander von Humboldt Foundation.

\bibliographystyle{asp2010}
\bibliography{betel}

\end{document}